\documentclass[showpacs, pra,onecolumn,preprintnumbers ,amsmath, amssymb, superscriptaddress, aps]{revtex4-2}
\usepackage{color}
\usepackage{amsmath,amssymb}
\usepackage{pifont}
\usepackage{amssymb}  
\usepackage{bbold}
\usepackage{float}
\usepackage{subfloat}
\usepackage[caption=false]{subfig}
\usepackage{tikz}
\usepackage{makecell}
\usepackage{subfig}
\usepackage{pifont}   
\usepackage{graphicx} 
\graphicspath{{Figures/}}
\usepackage{dcolumn}  
\usepackage{bm}       
\usepackage{multirow} 
\usepackage{placeins}
\usepackage[colorlinks]{hyperref}
\usepackage{mathtools}
\usepackage{appendix}

\begin{document}
	\title{{{Tuned gap in graphene through laser barrier}}}
	\date{\today}
		\author{Hasna Chnafa}
	\affiliation{Laboratory of Theoretical Physics, Faculty of Sciences, Choua\"ib Doukkali University, PO Box 20, 24000 El Jadida, Morocco}
		\author{Miloud Mekkaoui}
	\affiliation{Laboratory of Theoretical Physics, Faculty of Sciences, Choua\"ib Doukkali University, PO Box 20, 24000 El Jadida, Morocco}
	\author{Ahmed Jellal}
	\email{a.jellal@ucd.ac.ma}
	\affiliation{Laboratory of Theoretical Physics, Faculty of Sciences, Choua\"ib Doukkali University, PO Box 20, 24000 El Jadida, Morocco}
	\affiliation{Canadian Quantum  Research Center,	204-3002 32 Ave Vernon,  BC V1T 2L7,  Canada}
		\author{Abdelhadi Bahaoui}
	\affiliation{Laboratory of Theoretical Physics, Faculty of Sciences, Choua\"ib Doukkali University, PO Box 20, 24000 El Jadida, Morocco}
	
\begin{abstract} 
We study the effect of the energy gap on the transmission of fermions in graphene exposed to linearly polarized light as a laser barrier. We determine the energy spectrum, apply boundary conditions at interfaces, and use the transfer matrix approach to obtain transmissions for all energy modes. We show that when the energy gap increases, the oscillations of transmissions decrease dramatically until they vanish entirely. However, when the barrier width varies, the oscillations become more significant and exhibit sharp peaks. By increasing the incident energy, the laser field suppresses the Fabry-Pérot resonance, and the transmissions move to the right when the energy gap is tuned.

	\end{abstract}	
	\maketitle

\section{Introduction}
Graphene is a single sheet of carbon atoms organized in a two-dimensional honeycomb arrangement \cite{ao2}, which 
 is regarded as a miracle material because of its exceptional physical properties. It has a high resistivity \cite{ao4}, a high thermal conductivity \cite{s1ia}, a Hall effect \cite{s5,s6,s7}, a Klein tunneling \cite{s11a,s11b}, a mechanical strength \cite{s11c}, and an electronic mobility that is $100$ times greater than the most efficient material \cite{ao3}. It is also extremely transparent and light, making it an intriguing material for a broad range of applications, including electronics \cite{ao4}, photonics \cite{ao5}, and other fields.
At weak energies, the behavior of massless particles in graphene can be described by the Dirac equation, which gives rise to a linear dispersion relation \cite{s1a}. Nevertheless, the lack of a band gap in graphene limits its use in specific kinds of electronic devices. There have been several approaches developed to open such a band gap, for example, doping graphene with atoms such as nitrogen \cite{gr7} or boron \cite{gr8,gr9}.  {Depositing graphene on substrates generates a staggered potential in the graphene layer. Practically, this means that the carbon atoms in sublattices A and B have different values for the energy levels, designated by $\Delta$ and $-\Delta$,  respectively \cite{gr9a,gr9b}. Consequently, an energy gap emerges between the conduction and valence bands in the form of a mass term, which will be carried by the $\sigma_z$ component of the Pauli matrices.} The use of hexagonal boron nitride (hBN) as substrate, which possesses a lattice constant that closely resembles that of graphene, could result in the creation of a small band gap around $30$ \text{meV} \cite{gr1}.
 Silicone carbide (SiC) is another substrate used to produce a band gap $260$ \text{meV} in graphene
 \cite{gry1}.
  Also, a band gap can be opened by engineering graphene under suitable deformations \cite{gr3,gr4,gr5,gr6},
  and experimentally a band gap $300$ \text{meV} can be attained by applying $1\%$ uniaxial strain to graphene \cite{gr3}.

{Floquet theory has grown enormously in recent years due to its ability to explain the electronic behavior of periodically driven quantum mechanical systems subjected to a transient periodic field \cite{gr14c,gr14d}. It should be emphasized that recent advances in optical and microwave physics, as well as technological advances in lasers and emerging applications of quantum optics in condensed matter, have facilitated direct experimental verification of these theoretical predictions. This has resulted in concrete applications in real optoelectronic devices \cite{gr14e,gr14f,gr14g,gr14h,gr14i}. Several studies have been  done, examining the application of Floquet theory to different materials  \cite{gr15A,gr15B}, notably nanotubes \cite{gr15C,gr15D}, graphene \cite{gr15G,gr15H,gr15I},  $\alpha-\text{T}_{3}$  \cite{gr15E,gr15F} and tilted  \cite{gr15IA,gr15IB} materials, and so on. It has been shown that the use of Floquet theory under the influence of periodic electromagnetic fields can provide information about the electron (or optical) dressed states and control the electronic \cite{gr15H,gr15IC,nano1,tilted1}, topological \cite{gr15ID,grI16,grI17,grI18} and transport \cite{gr15IE,gr15IZ,gr16,gr17,Nano,Tilted} properties of these materials, opening up new potential applications in quantum electronics and spintronics fields.}

The study of laser effects on graphene-based systems is a well-established research field in modern physics. 
{Lately, it has been demonstrated that the laser field is applied to tune the electronic properties of graphene and create controllable and dynamic gaps in the quasi-energy spectrum based on the intensity and polarization of the field \cite{gr10,gr11,gr12,gr13,gr14,gr14a}. For circularly polarized light, it becomes possible to produce a band-gap at the Dirac point and generate dynamic gaps at various momenta. Moreover, these gaps can be adjusted by manipulating the field intensity \cite{gr14b}. In contrast, when linearly polarized light interacts with graphene, it affects its electrons in different ways, such as by absorbing or emitting photon energy. As a result, the energy levels of the electrons in graphene can vary depending on the direction of linear polarization, forming an anisotropic quasi-energy spectrum that displays dynamic gaps only in specific directions at non-zero momentum, while no gap is observed at the Dirac point. Thus, light can be considered an optimal tool for regulating the transport properties of graphene systems.
Furthermore, it is found that a strong monochromatic field induces asymmetric Klein tunneling at normal incidence \cite{gr15,gr16,gr17,gr18,gr18a,gr18b}. Indeed, when the laser field is present, the Hamiltonian takes on a non-commutative form with a chirality operator, and the resulting broken chirality leads to the loss of the symmetric Klein phenomenon  \cite{gr17}. Other results seen in the presence of a vector potential oscillating in time include the fact that the irradiation area of the graphene sheet acts as a barrier, making it likely to confine the Dirac fermions \cite{gr16}, the transmission via the electrostatic barrier in irradiated graphene can display the oscillations of the Fano kind \cite{gr17,grr17}, and the electron conductance can exhibit specific patterns by adjusting the laser field parameters \cite{gr17a}. One of the effects observed is the significant increase in conductance when the graphene is illuminated by the laser. Another important effect is the modulation of electronic conductance, where the conductance varies periodically.} Even though there have been several works on the subject, we think that tuning an energy gap in graphene under linearly polarized laser light could be of interest.


 
%
%

We study the effect of the energy gap on the transmission of the Dirac fermions in graphene subjected to a laser field of amplitude $F$ and frequency $\omega$ that acts a as barrier of height and width. We apply the Floquet approach and solve the Dirac equation to derive the eigenspinors, satisfying the continuity at interfaces. By using the matrix transfer method, we explicitly determine the transmission probabilities for 
energy modes $\varepsilon+m\omega$ ($m=0,\pm 1, \pm 2{,\cdots}$) \cite{6}.
By restricting ourselves to the central band $m=0$ and two sidebands $m=\pm1$, we show that the amplitudes and peaks of transmissions are strongly affected by the energy gap. Indeed, it is found that the transmissions over all channels exhibit oscillations with an amplitude that decreases gradually for a small energy gap and vanishes completely for a large one. This behavior changes with the increase in barrier width, resulting in more oscillations with sharp peaks. Furthermore, the Fabry-Pérot resonance is found to be suppressed when incident energy is increased.


The paper is organized as follows. In Sec. \ref{TFor}, we set a theoretical formalism including a Hamiltonian describing gapped graphene subjected to a time-dependent vector potential.  
Subsequently, we solve the eigenvalue equation and use the Floquet theory to get the solutions of energy spectrum.
These solutions at boundary conditions, together with the transfer matrix method, can be used to determine the transmissions associated with energy modes in Sec. \ref{TFSor}. To provide a better understanding, we numerically analyze and discuss  transmissions for the central band and  the first sidebands  in Sec. \ref{TFSor1}. We summarize  our work in Sec. \ref{TFSo}.

\section{Theoretical formulation}\label{TFor}
	
\begin{figure}[ht]
	\centering
	\includegraphics[width=10cm, height=5cm]{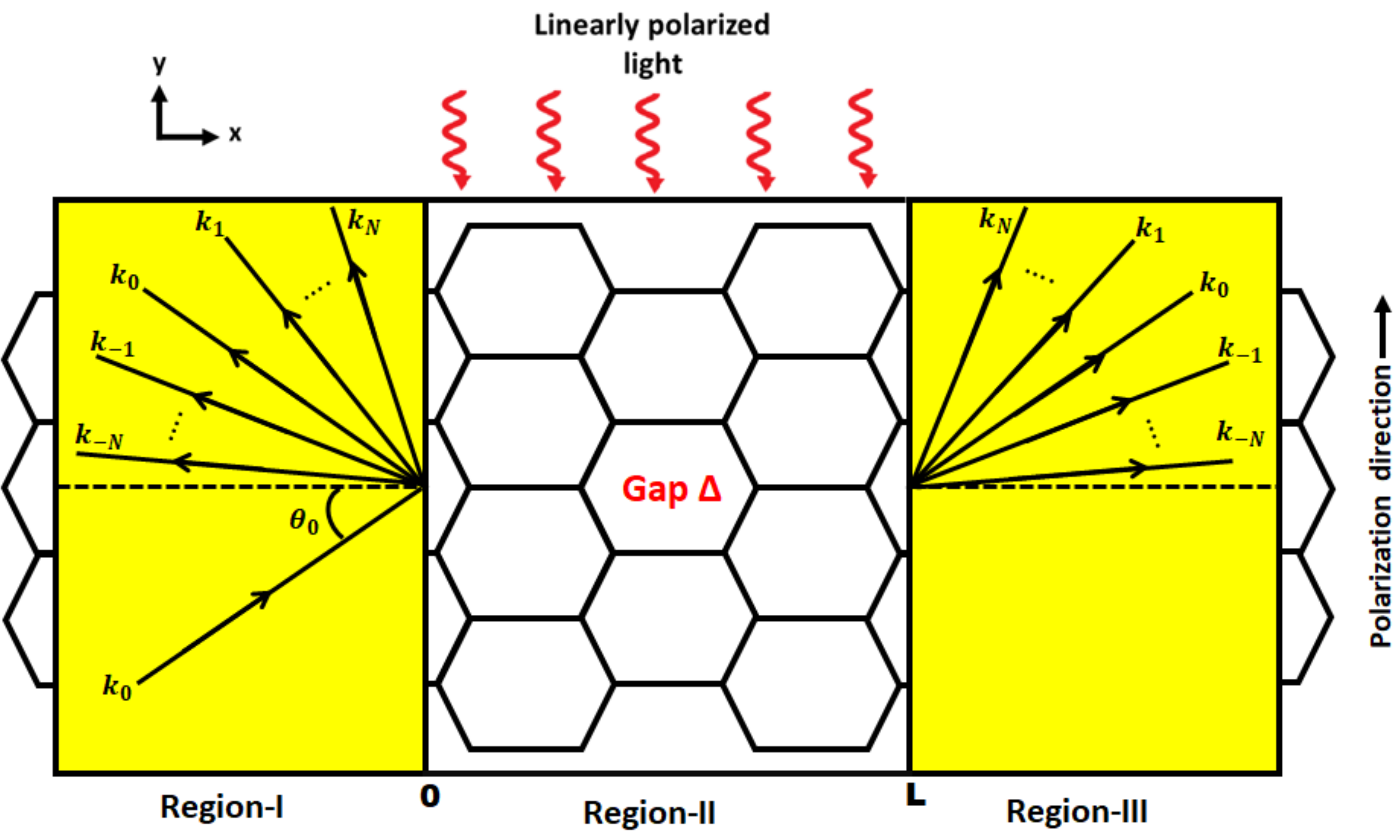}
	\caption{{(color online) Schematic illustration of gapped graphene exposed to a linearly polarized laser field.}}\label{db.5}
\end{figure}

We study the effect of energy gap engineering on transmission in a laser barrier made of gapped graphene subjected to a time-varying vector potential. This system is composed of regions labeled $\text{j}=\text{I},\text{II},\text{III}$ with a tuned gap $\Delta$ and a linearly polarized light applied to \text{II}, as depicted in  Fig.\ref{db.5}.

{The present system can be described by the following Hamiltonian}
\begin{align}\label{H1}
	\mathcal{H}=v_{F}\boldsymbol{\sigma}\cdot\left[\boldsymbol{p}-e\boldsymbol{A}\right]+{\Delta}\sigma_z
\end{align}
where $v_F\approx10^{6}$ \text{m/s} is the Fermi velocity, $\boldsymbol{p}=(p_x,p_y)$ is the momentum, $e$ is the electron charge, and $\Delta$ is the energy gap.
In the dipole approximation \cite{1}, the laser light is represented by the vector potential 
$	\boldsymbol{A}=\left(0,A\cos{\omega t}\right)$
 with $\omega$ is the frequency of a wave, and  the corresponding laser field is
$	\boldsymbol{E}=-\dfrac{\partial}{\partial t}\boldsymbol{A}=\left(0,F\sin{\omega t}\right)$,
$F={\omega}{A}$ with $F$ being the amplitude.

According to the Floquet approach  \cite{aq,aq1,aq3}, we write
the eigenspinors $\psi_{\text{j}}(x,y,t)
	=\left[\psi_{\text{A},\text{j}}(x,y,t),\psi_{\text{B},\text{j}}(x,y,t)\right]^{\text{T}}=e^{-i\mathcal{\varepsilon}t}\Phi_{\text{j}}(x,y,t)$ and therefore the eigenvalue equation gives
{\begin{align}\label{re}	\begin{pmatrix}
		\Delta&  -i{\partial_x}-{\partial_y}+i{\dfrac{F}{\omega}}\cos{\omega t} \\
		-i{\partial_x}+{\partial_y}-i{\dfrac{F}{\omega}}\cos{\omega t}  & -\Delta
	\end{pmatrix}\begin{pmatrix}\psi_{\text{A},\text{j}}(x,y,t)\\\psi_{\text{B},\text{j}}(x,y,t)\end{pmatrix}=i\partial_t\begin{pmatrix}\psi_{\text{A},\text{j}}(x,y,t)\\\psi_{\text{B},\text{j}}(x,y,t)\end{pmatrix}
\end{align} 
{where the dimensionless quantities are defined  $F\equiv{F}/{{F_0}}$, $\textbf{r}\equiv{\textbf{r}}/{L_{0}}$,  $\mathcal{\varepsilon}\equiv{\mathcal{\varepsilon}}/{\mathcal{\varepsilon}_{0}}$, ${\Delta}\equiv{\Delta}/{\mathcal{\varepsilon}_{0}}$, $t\equiv{t}/{t_{0}}$, $\omega\equiv{\omega}/{\omega_{0}}$ with scales:  length $L_{0}=\sqrt{\dfrac{\hbar v_{F}}{eF_{0}}}$, time  $t_{0}=\dfrac{L_{0}}{v_{F}}$, energy  $\mathcal{\varepsilon}_{0}=\dfrac{\hbar {\tiny }v_{F}}{L_{0}}$, frequency  $\omega_{0}=\dfrac{v_{F}}{L_{0}}$ and  electric field  $F_0=\dfrac{\mathcal{\varepsilon}_{0}}{eL_0}$.}
By taking into account the conservation of  $p_y$ and using the periodicity   $\Phi_{\text{j}}(x,y,t+2\pi/\omega)=\Phi_{\text{j}}(x,y,t)$, we can write the Fourier series \cite{gr16,3A}
\begin{align}\label{g1}
	\Phi_{\text{j}}(x,y,t)=e^{ik_{y}y}\Phi_{\text{j}}(x)e^{-i\frac{F}{\omega^{2}}\sin{\omega t}}=e^{ik_{y}y}\Phi_{\text{j}}(x)\sum^{
		+\infty}_{n=-\infty}J_{n}\left(\frac{F}{\omega^{2}}\right)e^{-i n\omega t}
\end{align}
where $J_{n}$ the first-order Bessel function of argument $\frac{F}{\omega^{2}}$. Note that a linear combination of spinors of energies $\varepsilon+m\omega$ ($m=0,\pm 1, \pm 2, \cdots$) should be considered since the laser field can change the amount of energy that electrons have in units of $\omega$. As a result, the solution of (\ref{re}) can be written as
\begin{align}\label{1}
	\psi_{\text{j}}(x,y,t)=e^{ik_{y}y}\sum^{
		+\infty}_{m,n=-\infty}\Phi^{m}_{\text{j}}(x)J_{n-m}\left(\frac{F}{\omega^{2}}\right)e^{-i \left(\mathcal{\varepsilon}+n\omega \right)t}.
\end{align}
As it is clearly seen, to get a complete solution, one has to determine $\Phi^{m}_{\text{j}}(x)$, needs to be done in the next step by considering each region separately.

In region-$\text{I}$ $(x<0)$, without  energy gap ($\Delta=0$) and  laser field ($F=0$),  the eigenspinors can be written by exchanging $J_{n-m}(0)$ by $\delta_{nm}$  
\begin{align}
	\psi_{\text{I}}(x,y,t)&={e^{ik_{y}y}}\sum^{+\infty}_{m,n=-\infty}\left[\delta_{m0}\begin{pmatrix}
		1\\
	\gamma_m
	\end{pmatrix}e^{i{k_{m}}x}
+ r_{m}  \begin{pmatrix}
	1\\
-\dfrac{1}{\gamma_m}
\end{pmatrix}e^{-i{k_{m}}x}\right]\delta_{nm}e^{-i\left(\mathcal{\varepsilon}+n\omega\right)t}
\end{align}
where $r_{m}$ is the reflection coefficient, the complex number  and the angle outside the barrier are given by
\begin{align}
\gamma_m=s_m\dfrac{k_{m}+ik_y}{\sqrt{k_{m}^{2}+k^{2}_{y}}}=s_m e^{i\theta_m},\quad \theta_m=\tan^{-1}\left(\frac{k_y}{k_{m}}\right)
\end{align}
with $s_m=\text{sgn}(\varepsilon+m\omega)$ is the sign function and we have set $k_{x,m}=k_m$. The corresponding energies and $x$-direction component of the wave vector read as
\begin{align}\label{a1}
&	{\mathcal{\varepsilon}+m\omega=s_m\sqrt{k_{m}^{2}+k^{2}_{y}}}
\\
&\label{Ar1}
	k_{m}=\sqrt{\left(\mathcal{\varepsilon}+m\omega\right)^{2}-k^{2}_{y}}.
\end{align}

In region-\text{II} ($0\leq x\leq L$), $F\neq 0$ and $\Delta\neq 0$, then we have 
\begin{align}\label{hy}
	\begin{pmatrix}
		-i\left(\varepsilon+m\omega{-\Delta}\right)	   &  \left(\partial_{x}+k_{y}-m\omega\right) \\
		\left(\partial_{x}-k_{y}+m\omega\right) &  	-i\left(\varepsilon+m\omega{+\Delta}\right)
	\end{pmatrix}\begin{pmatrix}
		\Phi^{m}_{\text{A},{\text{II}}}(x)\\
		\Phi^{m}_{\text{B},{\text{II}}}(x)
	\end{pmatrix}=\begin{pmatrix} 0\\0
	\end{pmatrix}
\end{align} 
which can be solved to end up with the eigenspinors
\begin{align}
	\Phi^{m}_{\text{II}}(x)&=\alpha_{m}\begin{pmatrix}
		\xi^{+}_{m}\\
		\xi^{-}_{m}\gamma'_{m}
	\end{pmatrix}e^{i{q_{m}}x}+ \beta_{m} \begin{pmatrix}
		\xi^{+}_{m}\\
		-\dfrac{\xi^{-}_{m}}{\gamma'_m}
	\end{pmatrix}e^{-i{q_{m}}x}
\end{align}
where we have set
\begin{align}
	\xi^{\pm}_{m}=\left(1\pm\dfrac{s'_m\Delta}{\sqrt{q_{m}^{2}+(k_{y}-m\omega)^{2}+\Delta^{2}}}\right)^{\frac{1}{2}}	
\end{align}
as well as the quantities
\begin{align}\label{mz7}
\gamma'_m=s'_m\dfrac{q_{m}+i\left(k_{y}-m\omega\right)}{\sqrt{{q_{m}^{2}}+(k_{y}-m\omega)^{2}}}=s'_m e^{i\theta'_m},\quad  \theta'_m=\tan^{-1}\left(\frac{k_y-m\omega}{q_{m}}\right)
\end{align}
with $s'_m=\text{sgn}(\varepsilon+m\omega)$, $q_{x,m}=q_m $, $\alpha_{m}$ and $\beta_{m}$ are two constants. 
We obtain the following eigenvalues and the wave component
\begin{align}\label{m7}
&	{\mathcal{\varepsilon}+m\omega=s'_m\sqrt{{q_{m}^{2}+(k_{y}-m\omega)^{2}{+\Delta^{2}}}}}
\\
&\label{A1}
		q_{m}=\sqrt{\left(\mathcal{\varepsilon}+m\omega\right)^{2}-(k_{y}-m\omega)^{2}{-\Delta^{2}}}.
\end{align}
Combining all to get the eigenspinors in region-\text{II} ($0\leq x\leq L$)
\begin{align}
	\psi_{\text{II}}(x,y,t)&={e^{ik_{y}y}}\sum^{
		+\infty}_{m,n=-\infty}\left[\alpha_{m}\begin{pmatrix}
		\xi^{+}_{m}\\
		\xi^{-}_{m}\gamma'_{m}
	\end{pmatrix}e^{i{q_{m}}x}+ \beta_{m} \begin{pmatrix}
		\xi^{+}_{m}\\
		-\dfrac{\xi^{-}_{m}}{\gamma'_m}
	\end{pmatrix}e^{-i{q_{m}}x}\right]J_{n-m}\left(\frac{F}{\omega^{2}}\right)e^{-i\left(\mathcal{\varepsilon}+n\omega\right)t}.
\end{align}

For region-\text{III} ($x>L$), the eigenspinors can be worked out as  in  region-\text{I} to get 
\begin{align}
	\psi_{\text{III}}(x,y,t)&={e^{ik_{y}y}}\sum^{+\infty}_{m,n=-\infty}\left[t_{m}\begin{pmatrix}
		1\\
    \gamma_m
	\end{pmatrix}e^{i{k_{m}}x}
	+ \lambda_m \begin{pmatrix}
		1\\
		-\dfrac{1}{\gamma_m}
	\end{pmatrix}e^{-i{k_{m}}x}\right]\delta_{nm}e^{-i\left(\mathcal{\varepsilon}+n\omega\right)t}
\end{align}
where $t_m$ is the transmission coefficient and $\lambda_m$ is the null vector. We will then look at how the results described above can be used to determine the transmission probabilities for all energy modes.

\section{Transmission Probabilities}\label{TFSor}

To determine the transmission probabilities in gapped graphene through a laser assisted barrier, we first introduce the continuity of eigenspinors at the interfaces $x=0,{L}$, i.e., ${\psi_{\text{I}}(0,y,t)}={\psi_{\text{II}}(0,y,t)}$ and ${\psi_{\text{II}}({L},y,t)}={\psi_{\text{III}}({L},y,t)}$. Second, 
by taking into account that $e^{in{\omega} t}$ as an orthogonal basis, then  at $x=0$ we have
{\begin{align}
	&	\delta_{n0}+r_{n} =\sum^{+\infty}_{m=-\infty}
		\left[\alpha_{m}\xi^{+}_{m}+\beta_{m}\xi^{+}_{m}\right]
		J_{n-m}{{\left(\frac{{F}}{{\omega}^{2}}\right)}} \label{eqx01}\\
	&	\delta_{n0}{\gamma_n}-{r_{n}} \dfrac{1}{\gamma_n} =
		\sum^{+\infty}_{m=-\infty}{\left[\alpha_{m}	\xi^{-}_{m}{\gamma'_{m}}-\frac{\beta_{m}	\xi^{-}_{m}}{{\gamma'_{m}}}\right]}
		J_{n-m}{{\left(\frac{{F}}{{\omega}^{2}}\right)}} \label{eqx02}
	\end{align}
and at $x=L$, we get
	\begin{align}
	&	t_{n}{e^{i{k_{n}}{L}}}+
		{\lambda_{n}}{e^{-i{k_{n}}{{L}}}} =\sum^{+\infty}_{m=-\infty} \left[\alpha_{m}\xi^{+}_{m}{e^{i{q_{m}}{L}}}+\beta_{m}\xi^{+}_{m}{e^{-i{q_{m}}{L}}}\right]J_{n-m} {{\left(\frac{{F}}{{\omega}^{2}}\right)}} \label{eqxd1} \\
	&	t_{n}{\gamma_n}{e^{i{k_{n}}{{L}}}}-{{\lambda_{n}}}\dfrac{1}{\gamma_n}{e^{-i{k_{n}}{L}}} 
		=\sum^{+\infty}_{m=-\infty}\left[\alpha_{m}\xi^{-}_{m}{\gamma'_{m}}{e^{i{q_{m}}{L}}}-\frac{\beta_{m}\xi^{-}_{m}}{{\gamma'_{m}}}{e^{-i{q_{m}}{L}}}\right]J_{n-m} {{\left(\frac{{F}}{{\omega}^{2}}\right)}}. \label{eqxd2}
\end{align}} 
These can be written in matrix form  as follows:
\begin{align}\label{ar1}
	\begin{pmatrix}
		{\mathbb{E}_{1}}\\
		{\mathbb{E}'_{1}}
	\end{pmatrix}={\mathbb V}\begin{pmatrix}
		\mathbb{E}_{2}\\
		\mathbb{E}'_{2}
	\end{pmatrix}
\end{align}
where the transfer matrix ${\mathbb V}$ is  given by
\begin{eqnarray}
	{\mathbb V}=\begin{pmatrix}
		{\mathbb{V}_{11}}   &{\mathbb{V}_{12}}\\
		{\mathbb{V}_{21}}&{\mathbb{V}_{22}}
	\end{pmatrix}={\mathbb V^{-1}_{1}}(0)\cdot{\mathbb V_{2}}(0)\cdot{\mathbb V^{-1}_{2}}(L)\cdot{\mathbb V_{1}}(0)\cdot{\mathbb V_{3}}(L)
\end{eqnarray}
and different matrices 
\begin{eqnarray}
	{\mathbb V_{1}}(0)=
	\begin{pmatrix}
		{\mathbb I}& {\mathbb I} \\
		{{\mathbb U^{+}}} &{{\mathbb U^{-}}} \\
	\end{pmatrix},  \quad {\mathbb V_{2}}(L)=
	\begin{pmatrix}
		{{\mathbb W^{+}(L)}} & {{\mathbb W^{-}(L)}} \\
		{{\mathbb Y^{+}(L)}} & {{\mathbb Y^{-}(L)}}
	\end{pmatrix}, \quad {\mathbb V_{3}}(L)=
	\begin{pmatrix}
		{{\mathbb Z^{+}(L)}}& {\mathbb O} \\
		{{\mathbb O}} &{{\mathbb Z^{-}(L)}} \\
	\end{pmatrix}
\end{eqnarray}
involving the set of elements
\begin{align} \label{eqn1}
&	\left({{\mathbb
			U^{\pm}}}\right)_{nm}=\pm\left({\gamma_{n}}\right)^{\pm
		1}\delta_{nm}\\ 
&	\left({{\mathbb
			W^{\pm}(L)}}\right)_{nm}=\xi^{+}_{m}e^{\pm iq_{m}{L}}J_{n-m}{{\left(\frac{{F}}{{\omega}^{2}}\right)}} \\ &\left({{\mathbb Y^{\pm}(L)}}\right)_{nm}=\pm \xi^{-}_{m} ({\gamma'_{m}})^{\pm 1}e^{\pm iq_{m}{L}}
	J_{n-m}{{\left(\frac{{F}}{{\omega}^{2}}\right)}}  \\ 	
&	\left({\mathbb Z^{\pm}(L)}\right)_{nm}=e^{\pm i{k_{n}}{L}}\delta_{nm} 
\end{align}
with ${\mathbb O}$ is the null matrix,  ${\mathbb I}$ is the unit matrix,  ${\mathbb{E}_{1}}=\{{\delta_{0m}}\}$, ${\mathbb{E}'_{1}}=\{r_{m}\}$,  $\mathbb{E}_{2}=\{t_{m}\}$ and $\mathbb{E}'_{2}=\{\lambda_{m}\}=0$. These allow us to establish the relation
\begin{align}\label{n1}
\mathbb{E}_{2}={\mathbb{V}_{11}^{-1} \cdot\mathbb{E}_{1}}.
\end{align} 
This involves several modes that cause a problem in determining the transmission coefficient $t_m$. To overcome such a situation, we truncate these infinite series into a limited number within the range of $-N$ to $N$ with $N>F/\omega^{2}$ \cite{5,6,7,8}. Consequently, we write
\begin{equation}\label{eqt}
	t_{-N+k}={\mathbb{V}'}\left[k+1, N+1\right], \quad k=0, 1,\cdots, 2N
\end{equation}
where {${\mathbb V^{'}}$} is the inverse matrix ${\mathbb{V}_{11}^{-1}}$.

To derive the transmission probabilities $T_{m}$, 
we use the transmitted ${\boldsymbol{J}_{\textbf{tran},m}}$ and incident ${\boldsymbol{J}_{\textbf{inc}, 0}}$ current densities, which can obtained from 
\begin{align}
\boldsymbol{J}=\psi^\dagger(x,y,t) \sigma_{x}\psi(x,y,t).
\end{align}
to derive the transmission probabilities  $ T_{m}=\left|\dfrac{{\boldsymbol{J}_{\textbf{tran},m}}}{{\boldsymbol{J}_{ \textbf{inc},0}}}\right|$
as follows 
\begin{align}\label{ma4}
		T_{m}
		=\dfrac{s_m}{s_0}\dfrac{\cos\theta_m}{\cos\theta_0}{\left|{t_{m}}\right|^{2}}.
\end{align}
In illustrating the above results and due to numerical complexity, we can limit ourselves to three transmission channels: the central band ($m=0$) and the two first sidebands ($m=\pm 1$)
\begin{align}\label{gt}
	t_{-1}={\mathbb V^{'}}[1,2], \quad t_{0}={\mathbb V^{'}}[2,2], \quad t_{1}={\mathbb V^{'}}[3,2].
\end{align}
Next, we will analyze and discuss the transmissions under various conditions to underline the basic features of the present system.

\section{Results and discussions}\label{TFSor1}
Fig. \ref{fi4} shows the transmission probabilities versus the energy gap $\Delta$ for $L=1.2$, $k_y=2$. In Fig. \ref{fi4}(\text{a}), we present the transmission for the static barrier $T_{s0}$ ($F/\omega^{2}=0$), central band $T_0$ and the two first sidebands $T_{\pm1}$ ($F/\omega^{2}=0.6$). We observe that $T_{s0}$ starts from unity (Klein tunneling) and exhibits oscillations when $\Delta<\varepsilon$, where the maxima of their amplitude remain constant at $T_{s0}=1$ (transmission mode), as seen also in \cite{az,azz}. It vanishes in the opposite case $\Delta>\varepsilon$ because the wave vector $q_{0}$ inside the barrier becomes imaginary, resulting in the emergence of evanescent states. 
By choosing $F/\omega^{2}=0.6$, we see that $T_0$ displays the same behavior as for $T_{s0}$ except that its magnitude drops sharply. Moreover, $T_{\pm1}$ is constant until $\Delta\simeq3$ after which it shows peaks in the zone $3\lesssim\Delta\lesssim16$ with different amplitudes and becomes null for larger $\Delta$. 
Contrary to a barrier oscillating in time \cite{8}, we notice that the transmission for single photon absorption and emission do not coincide at lower energy gaps. Fig. \ref{fi4}(\text{b}) presents the transmission for no photon exchange $T_0$ versus $\Delta$ with three values of incident energy $\varepsilon=15$ (blue line), $\varepsilon=20$ (green line), and $\varepsilon=25$ (red line). 
By increasing  $\varepsilon$, we see that $T_0$ starts to fluctuate from different points and moves to the right, as well as that the number of its oscillations augments rapidly. It is also worth noting that the amplitude of transmission reduces dramatically until it disappears at $\Delta>\varepsilon$ corresponding to an evanescent mode. These results show that $\Delta$ affects the transmissions, and then it can be used to tune the tunneling effect.

\begin{figure}[H]\centering
	\subfloat[]{\includegraphics[width=0.5\linewidth, height=0.19\textheight]{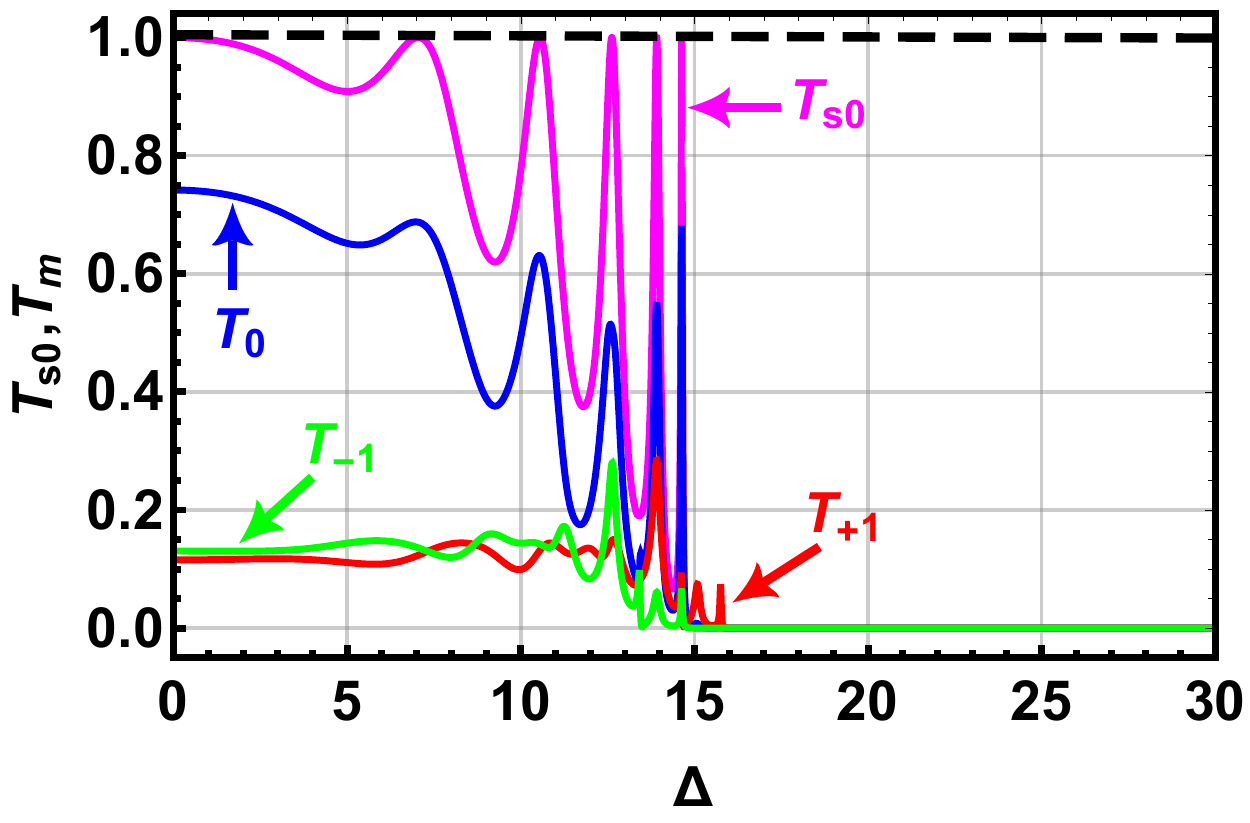}\label{d}}\subfloat[]{\includegraphics[width=0.5\linewidth, height=0.19\textheight]{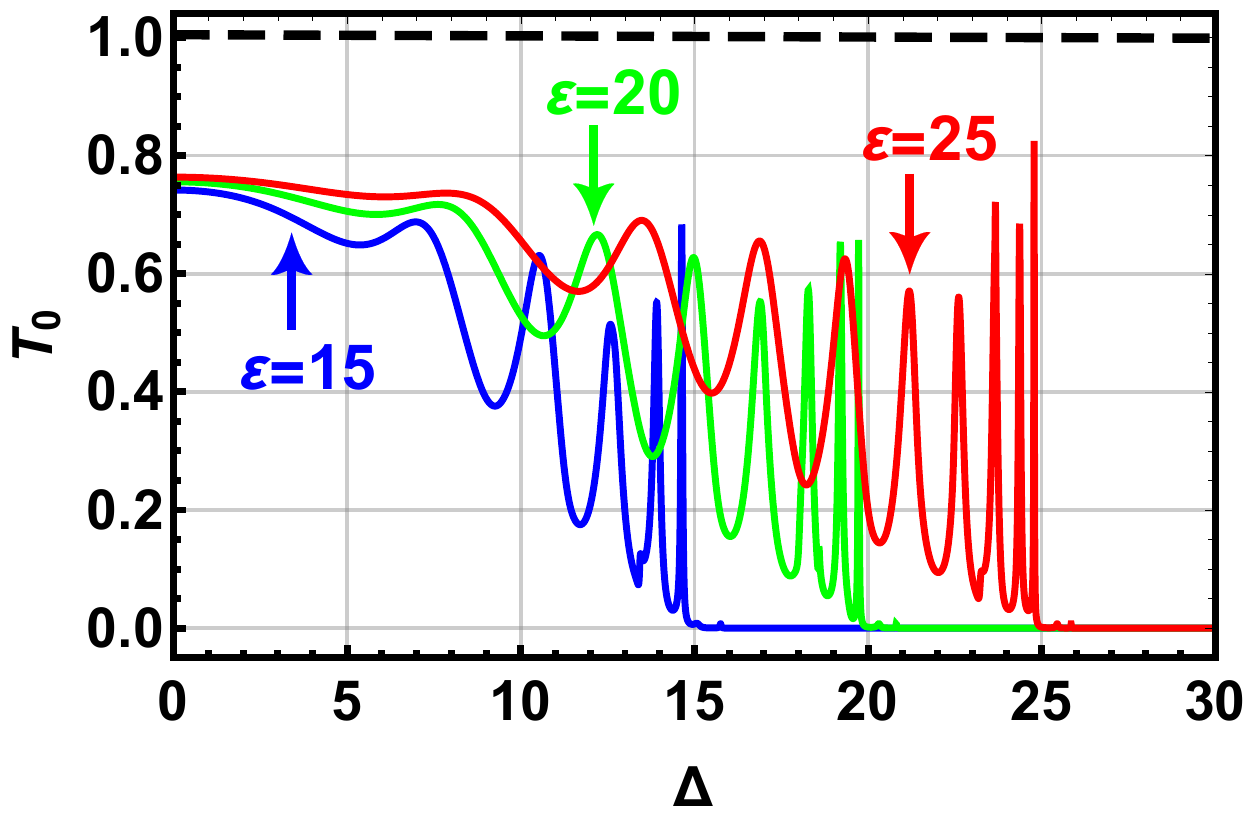}\label{dhu}}
	\caption{{{{(color online)  Transmission probabilities versus the energy gap $\Delta$ for ${L}=1.2$, $k_y=2$. \text{{(a)}}\color{black}{:} $\varepsilon=15$, $F=0$, $\omega=0$ for static barrier $T_{s0}$ (magenta line), and $F=0.6$, $\omega=1$ for central band $T_{0}$ (blue line) and two first sidebands $T_{-1}$ (green line), $T_{+1}$ (red line)}.  \text{{(b)}}\color{black}{:}  Transmission probability for the central band $T_0$ with $\varepsilon=15$ (blue line), $\varepsilon=20$ (green line), $\varepsilon=25$ (red line)}.}}\label{fi4}
\end{figure}

 Fig. \ref{fiA4} shows  the effect of amplitude $F$ and frequency $\omega$ of the laser field on the transmission $T_0$ as a function of  $\Delta$  for ${L}=1.2$, $k_y=2$, $\varepsilon=15$. In Fig. \ref{fiA4}\text{(a)}, we choose  $\{\omega=1, F=0.3,0.6,0.9\}$  and observe that $T_0$ decreases smoothly when  $F$ increases. 
 By fixing  $F$ and varying {$\omega$} as shown in Fig. \ref{fiA4}\text{(b)} with $\{F=0.6, \omega=1,1.3,2.6\}$, we notice that $T_0$ increases rapidly by increasing $\omega$ until the maxima of its oscillation become constant at $T_0=1$ (red line), which is in agreement with the results obtained in \cite{3A,3d}. Consequentially, the sharp peaks of the central band are greatly influenced by the laser field.

\begin{figure}[H]\centering
	\subfloat[]{\includegraphics[width=0.5\linewidth, height=0.19\textheight]{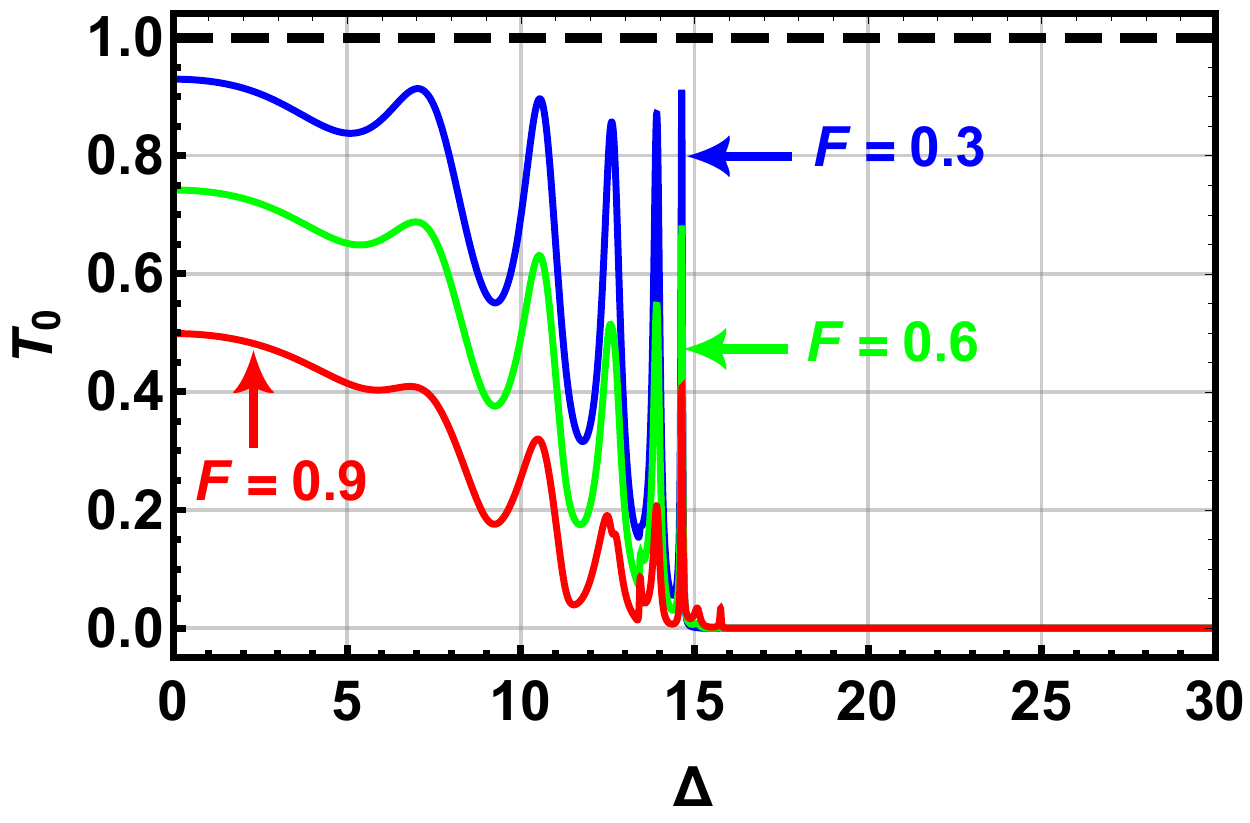}\label{1d}}\subfloat[]{\includegraphics[width=0.5\linewidth, height=0.19\textheight]{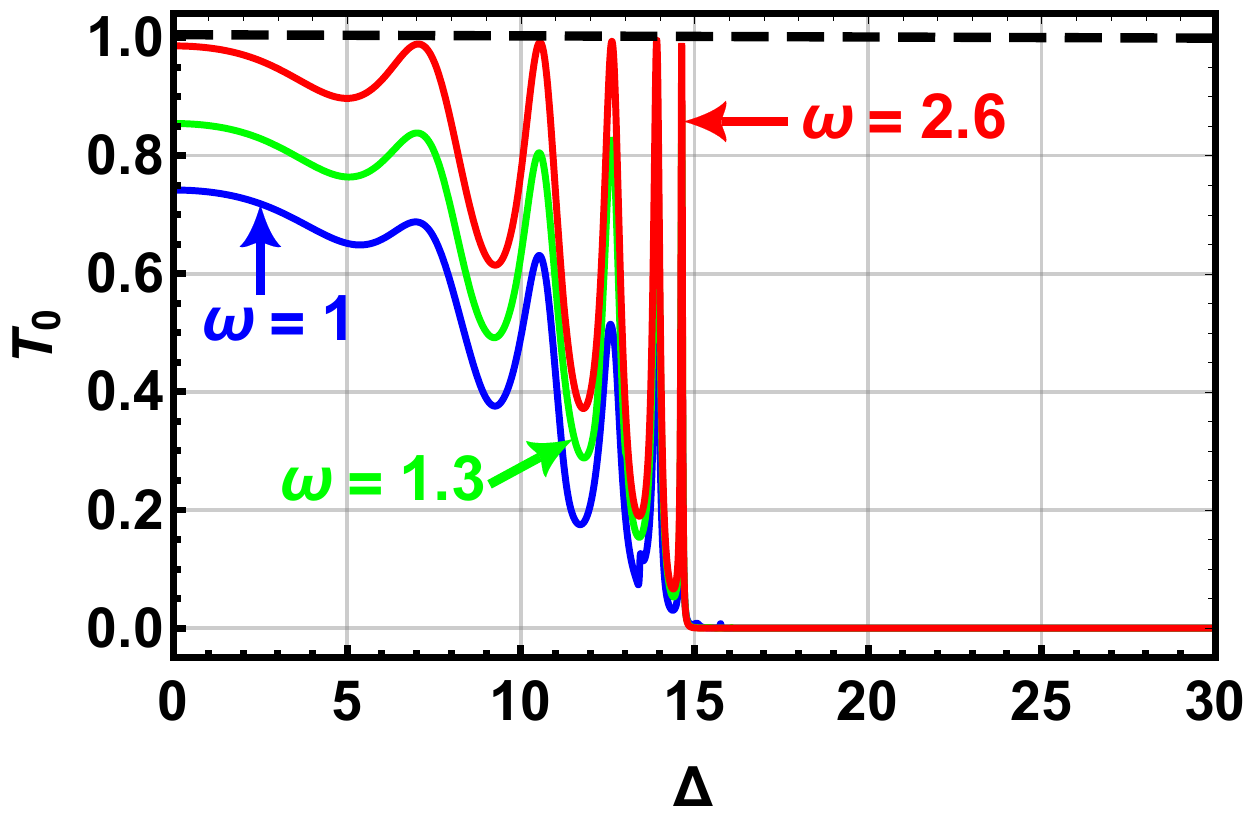}\label{1du}}
	\caption{{{ (color online)  Transmission probability for the central band $T_0$ versus the energy gap $\Delta$ for ${L}=1.2$, $k_y=2$, $\varepsilon=15$. \text{{(a)}}\color{black}{:} $\omega=1$, $F=0.3$ (blue line), $F=0.6$ (green line), $F=0.9$ (red line)}. \text{{{(b)}}}\color{black}{:} $F=0.6$, $\omega=1$ (blue line), $\omega=1.3$ (green line), $\omega=2.6$ (red line).}}\label{fiA4}
\end{figure}

The transmission probabilities of a static barrier $T_{s0}$ ($F/\omega^{2}=0$) and three first sidebands $T_m$ ($F/\omega^{2}=0.6$) versus the barrier width $L$ are shown in Fig. \ref{f4} for $k_y=2$, $\varepsilon=15$ and $\Delta=0,7$.
It is clearly seen  that in Fig. \ref{f4}\text{(a)} $T_{s0}$ is unity, and there is a perfect transmission as noted in \cite{zeb}. 
For $F/\omega^{2}=0.6$, we observe that the transmission $T_0$ of the central band initiates at unity and periodically oscillates, which it reduces under the increase of $L$ as found in \cite{3B}, contrary to a scenario in which the barrier oscillates over time and the maximum value of the peaks remains constant at unity \cite{7}. 
However, the transmissions ${T_{\pm1}}$ for the two sidebands {start} at zero and {exhibit} a sinusoidal shape. For $\Delta\neq0$ in Fig. \ref{f4}\text{(b)}, we see that $T_{s0}$ oscillates with an unchanging amplitude as long as $L$ increases. 
The amplitudes of the transmissions $T_m$ change dramatically and become much more oscillatory. The observation that ${T_{\pm1}}$ oscillate less than $T _0$ of the central band is intriguing.
We see an increase in the total number of peaks for all modes, a steady decrease in the $T_0$ maxima, and a downward shift of the $L$-dependent minima. These are a consequence of the presence of an energy gap $\Delta$ and laser field in the intermediate region of the present system.

\begin{figure}[H]\centering
	\subfloat[]{\includegraphics[width=0.5\linewidth, height=0.19\textheight]{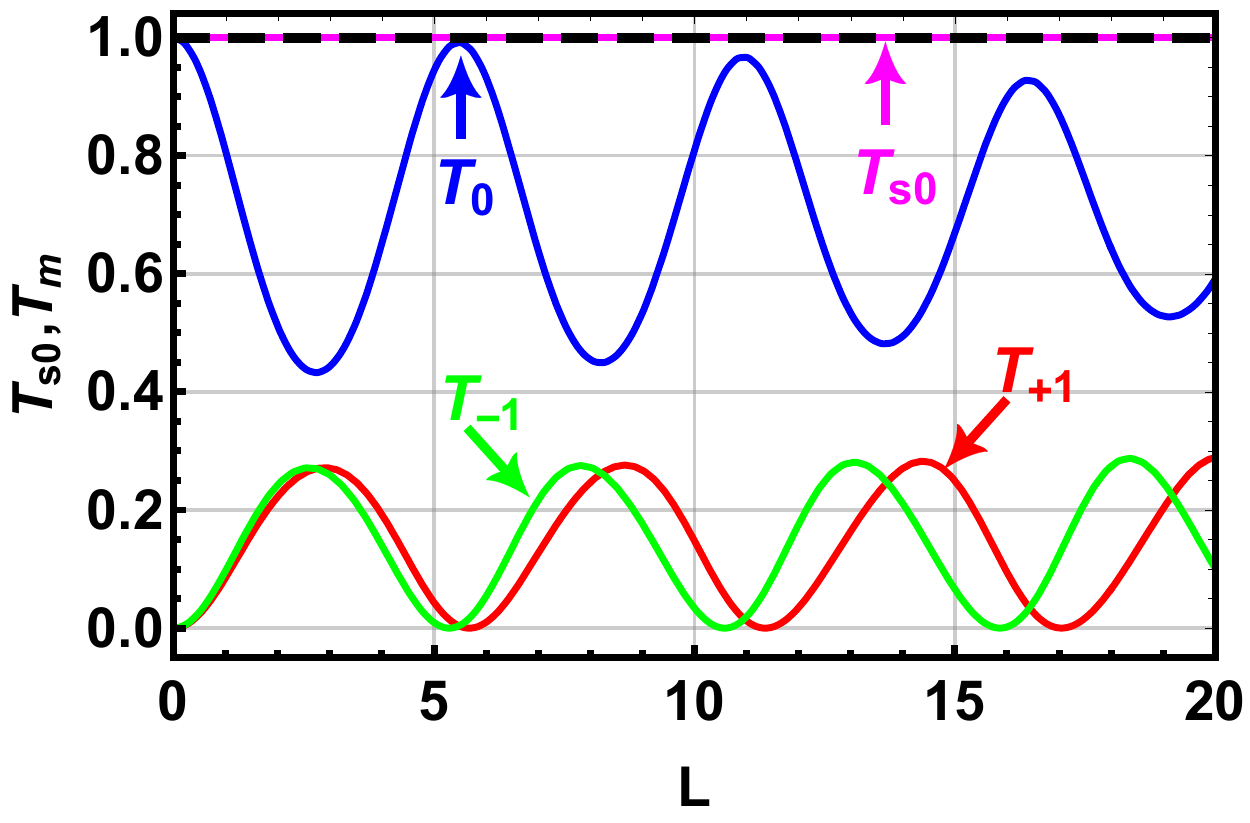}\label{1u}}\subfloat[]{\includegraphics[width=0.5\linewidth, height=0.19\textheight]{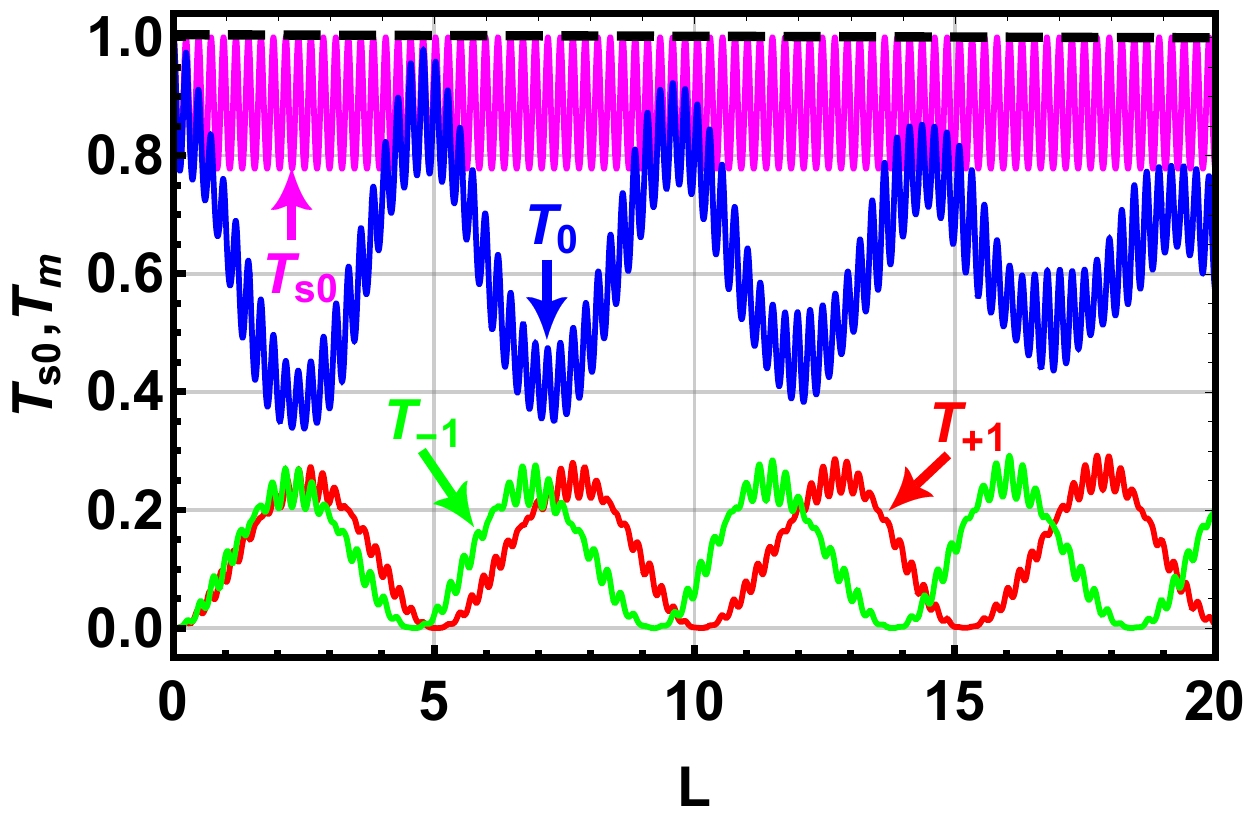}\label{1ud}}
	\caption{{{ {(color online)  Transmission probabilities versus the barrier width $L$ for  $k_y=2$, $\varepsilon=15$, $F=0$, $\omega=0$  for static barrier $T_{s0}$ (magenta line), and $F=0.6$, $\omega=1$ for central band $T_{0}$ (blue line) and two first sidebands $T_{-1}$ (green line), $T_{+1}$ (red line)}.  \text{{(a)}}\color{black}{:}   $\Delta=0$. \text{{(b)}}\color{black}{:}   $\Delta=7$.} }}\label{f4}
\end{figure}

In Fig. \ref{fz4}, we present the transmission  $T_{-1}$ versus the barrier width $L$. We choose $k_y=2$, $\varepsilon=15$, $\Delta=7$ with $\omega=1$, $F=0.25$ (blue line), $F=0.45$ (green line), $F=0.65$ (red line) in Fig. \ref{fz4}\text{(a)} and  $F=0.6$, $\omega=1$ (blue line), $\omega=1.15$ (green line), $\omega=1.3$ (red line) in Fig. \ref{fz4}\text{(b)}. We observe that $T_{-1}$ exhibits oscillating behavior with sharp peaks as a result of $\Delta\neq 0$. {This kind of oscillation resembles a plane wave modulated by higher frequency \cite{mod}, which has been attributed to the presence of the Bessel function in the transmission expansion. Physically, it can be attributed to constructive or destructive quantum interference phenomena between the transmitted and reflected electron waves. More specifically, when the width of the barrier is altered, the electron waves propagate across the barrier and interact with each other. This can lead to enhancements or cancellations of the transmission amplitude, resulting in periodic oscillations.} 
In Fig. \ref{fz4}\text{(a)}, we observe that the amplitude of $T_{-1}$ increases slowly with the increase of  $F$, as seen in \cite{3A}. 
Now we plot $T_{-1}$ in Fig. \ref{fz4}\text{(b)} by fixing $F$ {and} varying $\omega$. We observe that the magnitude of oscillations decreases quickly, and its number increases, shifting to the left when $L$ increases \cite{3A}.

\begin{figure}[H]\centering
	\subfloat[]{\includegraphics[width=0.5\linewidth, height=0.19\textheight]{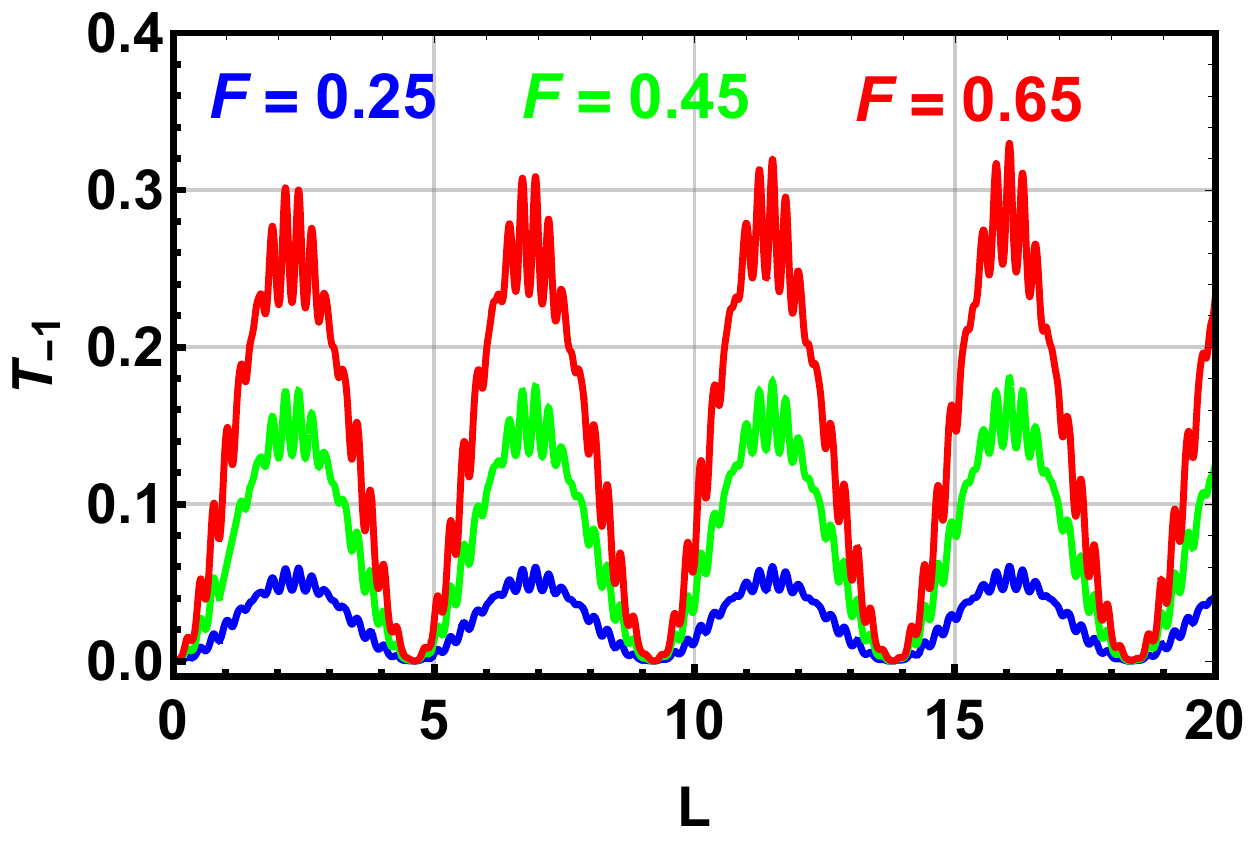}\label{zu}}\subfloat[]{\includegraphics[width=0.5\linewidth, height=0.19\textheight]{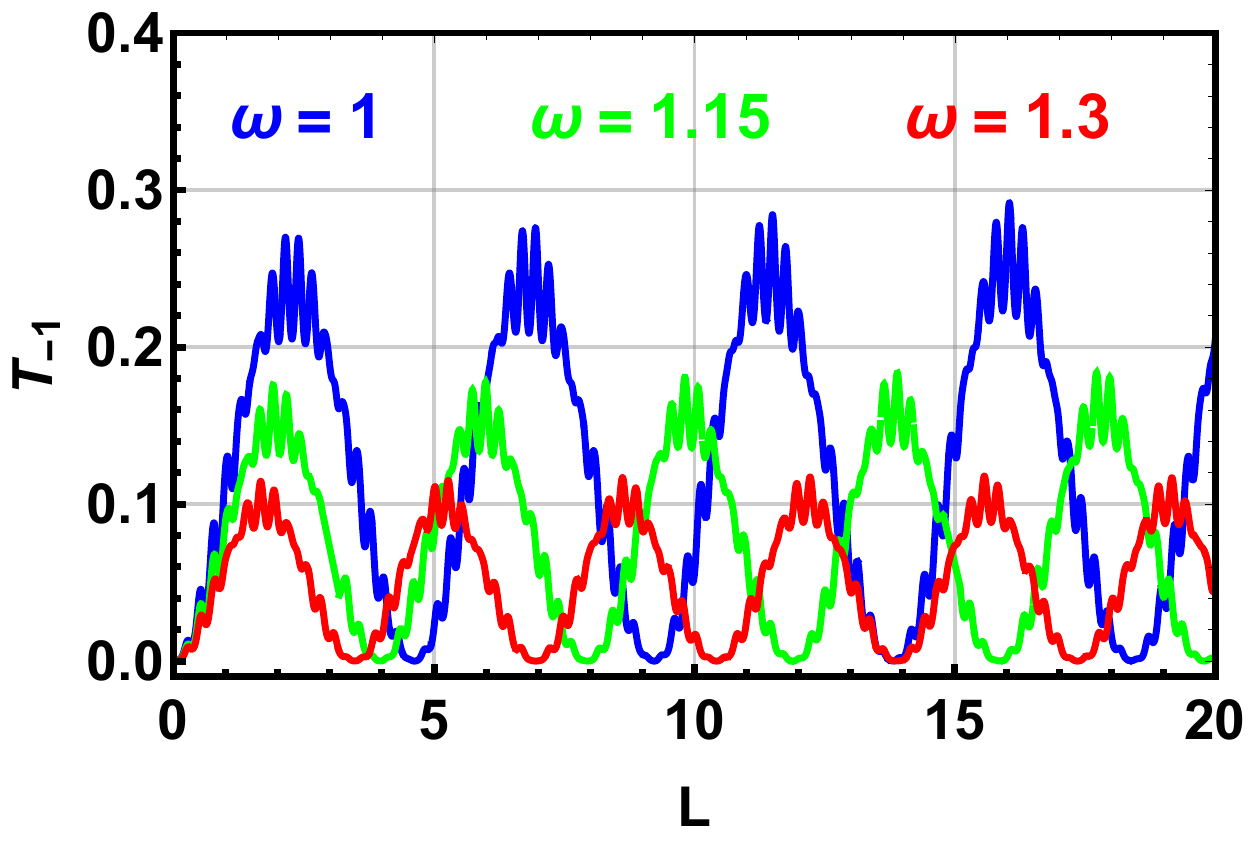}\label{zd}}
	\caption{{{{(color online)  Transmission probability for single photon emission $T_{-1}$ versus the barrier width $L$ for $k_y=2$, $\varepsilon=15$, $\Delta=7$. \text{{(a)}}\color{black}{:} $\omega=1$, $F=0.25$ (blue line), $F=0.45$ (green line), $F=0.65$ (red line)}. \text{{{(b)}}}\color{black}{:} $F=0.6$,   $\omega=1$ (blue line), $\omega=1.15$ (green line), $\omega=1.3$ (red line).} }}\label{fz4}
\end{figure} 

Fig. \ref{f} shows the effect of  $\Delta$ on the transmission probabilities versus the incident energy $\varepsilon$ with ${L}=1.2$, $k_y=2$. 
For $\Delta=7$ in Fig. \ref{f}(a), we observe that the transmission of the static barrier (magenta line) is perfectly zero when $\varepsilon\leq7$, i.e., total reflection. It exhibits the Fabry-Pérot oscillations, which appear at $\varepsilon\geq7$, while its amplitude almost disappears, nearing ($T_{s0}\sim1$) for larger $\varepsilon$ as seen in \cite{grr17,grr19}. 
When the laser field is activated, we clearly see that $T_0$ is more dominant than ${T_{\pm1}}$ and displays the same behavior as $T_{s0}$, with the exception that the maxima of its amplitude become less than one, as shown in \cite{grr17}. 
{Furthermore, we notice that ${T_{\pm1}}$ are null for small $\varepsilon$, which occurs because the charge carriers cannot cross the barrier due to the presence of a forbidden region where no electronic states are allowed in the conduction or valence band. Afterwards, they  display several peaks with variant amplitudes in the region $6\leq\varepsilon\leq30$ and become nearly linear, especially at larger $\varepsilon$.  This can be explained by noting that as the incident energy of the electrons increases, some electrons acquire sufficient energy to cross the energy gap and enter the conduction band. This leads to a decrease in the probability of transmission through graphene, as some of the energy is used to excite the electrons above the gap and create electron-hole pairs. As a result, transmission oscillations decrease in amplitude.}
Fig. \ref{f}(b), we depict ${T_{+1}}$ versus  $\varepsilon$ for $\Delta=6$ (blue line), $\Delta=7$ (green line), $\Delta=8$ (red line). 
One sees that the amplitude and shape of $T_{+1}$ are affected by $\Delta$, because the maxima of oscillations increase as $\Delta$ increases and decrease for larger $\varepsilon$.  {Another observation discovered is the presence of an asymmetric Fano resonance in the transmission \cite{Fano1, Fano2} around $\varepsilon=10$. These peaks may be attributed to quantum interference phenomena between discrete and continuous states that arise here due to the interaction between the laser field and the Dirac fermions.  The interferences can be either destructive, leading to a decrease in transmission amplitude, or constructive, resulting in an increase in transmission amplitude, as shown in Fig. \ref{f}(b).} 

\begin{figure}[H]\centering
	\subfloat[]{\includegraphics[width=0.5\linewidth, height=0.19\textheight]{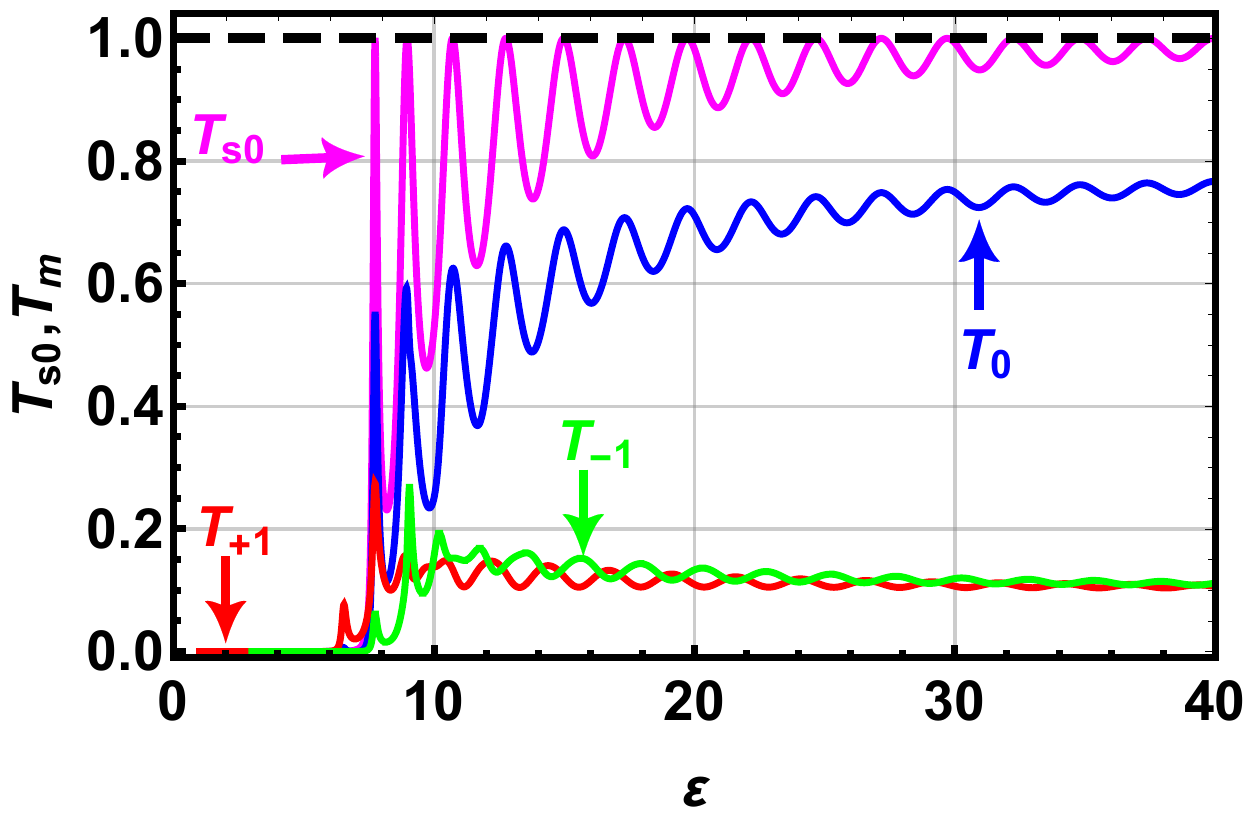}\label{zo}}\subfloat[]{\includegraphics[width=0.5\linewidth, height=0.19\textheight]{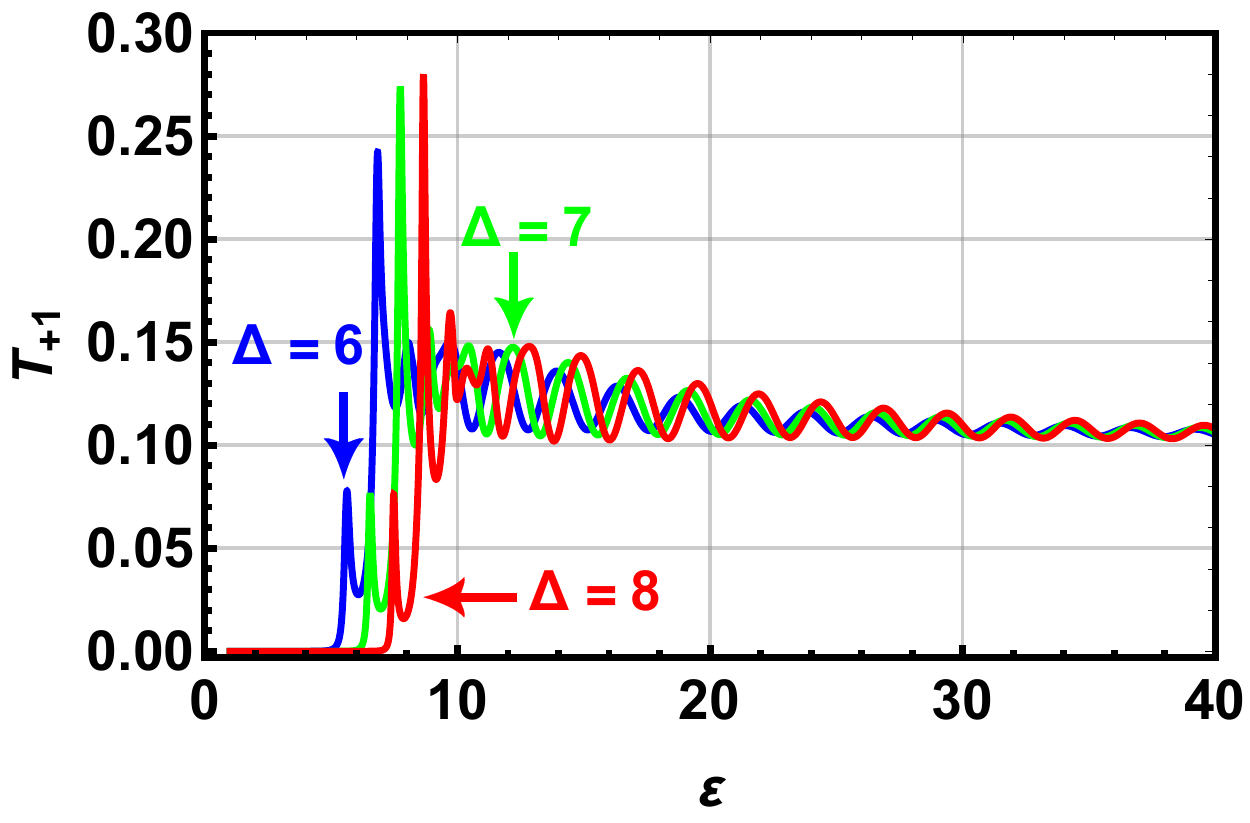}\label{md}}
\caption{{{{(color online) Transmission probabilities versus the incident energy $\varepsilon$ for ${L}=1.2$, $k_y=2$. \text{{(a)}}\color{black}{:} $\Delta=7$, $F=0$, $\omega=0$ for static barrier $T_{s0}$ (magenta line), and $F=0.6$, $\omega=1$ for central band $T_{0}$ (blue line) and two first sidebands $T_{-1}$ (green line), $T_{+1}$ (red line)}.  \text{{(b)}}\color{black}{:}  $T_{+1}$ with $\Delta=6$ (blue line), $\Delta=7$ (green line), $\Delta=8$ (red line)}.}}\label{f}
\end{figure}

\section{Conclusion}\label{TFSo}

We have studied the effect of an energy gap $\Delta$ on the tunneling transport of Dirac fermions in graphene subjected to a linearly polarized laser field of amplitude $F$ and frequency $\omega$. The eigenvalues and eigenvectors were obtained by applying the Floquet method and solving the Dirac equation in three regions composing the present system.
These have been merged at interfaces, and by using the transfer matrix method and current densities  transmission probabilities for all energy modes have been established. For simplicity and a better understanding,  we have restricted ourselves to  three transmission modes: central band ($m=0$), first sideband of photon absorption ($m=+1$) and first sideband of photon emission ($m=-1$).

Subsequently, we have numerically analyzed the transmission probabilities $T_m$ in appropriate conditions of  $\Delta$,  $F$, $\omega$, barrier width $L$ and incident energy $\varepsilon$. 
It has been found that the transmissions exhibit oscillations that gradually decrease at $\varepsilon<\Delta$ and disappear entirely at $\varepsilon>\Delta$, which corresponds to an evanescent mode.
 We have found that the transmissions of photon absorption $T_{+1}$ and emission $T_{-1}$ do  not meet for a small $\Delta$ as opposed to a barrier that oscillates in time, where $T_{+1}$ and $T_{-1}$ are identical. 
We have shown that $F$ and $\omega$ have a considerable effect on the behavior of  the transmission $T_m$ $(m=0,\pm1)$. Indeed, they reduce or grow slightly as long as $F$ varies, and they also rise or drop drastically with the change of $\omega$.  
 Another important finding with $\Delta$ is that the transmissions become significantly more oscillatory as the barrier width increases, in addition to an increase in the number of oscillations. Moreover, their maxima also show resonance that resembles sharp peaks.
  Furthermore, we have discovered that by changing the incident energy the laser field suppresses the Fabry-Pérot oscillations and the transmissions $T_m$ shift to the right as $\Delta$ varies. 
  
 On the basis of the experiment \cite{am} showing that the oxidation in irradiation regions via a two-photon mechanism results in the production of a band gap in the graphene, our conclusions can also be confirmed experimentally. {It is important to note that the energy gap can be tuned via several parameters, such as the amplitude of the laser field and its frequency, the barrier width, and the incident electron energy. The findings presented in this study demonstrate a novel approach to tuning the energy gap of graphene, offering exciting possibilities to generate optoelectronic devices based on graphene for technological applications in the future.}

\end{document}